# Paratransit Agency Responses to the Adoption of Sub-contracted Services Using Secure Technologies

Amari N. Lewis and Amelia C. Regan
Department of Computer Science
And Institution of Transportation Studies
University of California, Irvine

## 1 Abstract

Transportation agencies across the United States have the responsibility of providing transportation services for all travelers. Paratransit services which are designed to meet the needs of disabled travelers have been available to a certain extent for decades, but under the Americans with Disabilities Act mandate of 1990, uniform requirements were adopted across U.S. agencies. Most of these paratransit operators offer services which must be scheduled at least a day in advance. And, provision of these services by accessible busses is generally very expensive. Therefore, many agencies are considering sub-contracting some services to approved ride-hailing or taxi services. The purpose of this work is to examine the opinions of various public agencies with respect to the adoption of sub-contracted services through the use of secure technologies. Our research provides insight into the future of these partnerships. Agencies expressed interest in the use of privacy preserving secure technologies as well as a strong desire for better software solutions for paratransit passengers and operators. The on-line survey received thirty responses for a completion rate of 19.1%. Our primary findings are that a major concern of agencies for this sort of arrangement is the lack of Wheelchair Accessible Vehicles offered by taxis and TNCs and about 36% of the surveyed agencies have not considered such partnerships.

## 1 Motivation

Paratransit services are essential to the well-being of the populations served, but, nearly thirty years after the ADA was passed, they remain notoriously expensive to provide and inconvenient for users. A potential solution for improving paratransit services under consideration by transit agencies around the world, is the integration of ride hailing companies (also known as Transportation Network Companies, or TNCs) and taxi services. Passengers qualify for paratransit services because they are unable to use traditional transit due to a disability. These disabilities can range from mild to severe, in the sense that some require accessible accommodations such as wheelchair ramps, while those with vision impairments or milder mobility disabilities may not require such accommodations.

In fact, for vision impaired travelers with guide dogs, the wheelchair enabled vehicles can be dangerous. Due to the extra space in the vehicles, the guide dogs are not securely placed and are subject to moving around the vehicle rather than being able to lie close to their owners. However, enabling these services at the same time as ensuring user privacy, data security and safety will require considerations not currently available in standard ride-hailing and taxi services. There are some options outside of paratransit including;

- Gogograndparent (Gogograndparent, 2019) "Use Lyft or Uber without a smartphone" was introduced in 2016 and provides a phone based service to match customers with pre-approved ride-hailing vehicles or taxis based on their individual needs. Available throughout the U.S. and Canada.

- Uber WAV (WAV, 2020) was introduced in 2015 as an accessible option for wheelchair passengers. WAV is currently pilot testing in Chicago, DC, New York City, Philadelphia, Boston, Los Angeles, San Francisco, Portland, Phoenix, Houston, Austin, Toronto, the UK, Bangalore, Paris, and Newcastle (Australia).

- Uber Assist (Uber, 2020) for senior persons and passengers with disabilities Uber Assist provides certified drivers to give special assistance to riders who may need extra help. Currently available in over 40 cities and limited countries.

- Lyft (Lyft, 2020) has also partnered with companies like AARP, Access2care and many more to provide the Non-Emergency Medical Transportation (NEMT) service introduced in 2017. The pilot is now in Arizona, and has expanded to 5 more states as of 2019. Lyft became the first national ridesharing company to bring its transportation solutions to millions of medicaid beneficiaries.

- Lyft ACCESS mode (Lyft, 2020)- this is an accessible vehicle dispatch option where access mode allows customers to request a vehicle with wheelchair accommodations. If it is not available, it will provide dispatch options for passengers in their area.

But, some of these services are in pilot modes (even after several years) and others are available but only in some cities as previously mentioned which limits the ability for passengers to benefit from these services.

This research began with an exploration of using blockchain technology, an increasingly popular tool for privacy preserving secure contracting in logistics and supply chain networks, to create HIPAA (U.S. Health Insurance Portability and Accountability Act) (Health and Services, 2019) compliant contracting services (Lewis and Regan, 2020). Several other researchers have been examining similar solutions (Kanza and Safra, 2018; Luo et al., 2018; Javaid, Naveed, and Biplab, 2019), but after careful consideration we believe that a much simpler secure database would be a better solution for most agencies.

The purpose of this phase of our research is to gain insight from agency experts. Transit professionals with up to 20 years of experience allowed us to obtain diverse perspectives from the industry, and included Paratransit services department managers, ADA compliance and program managers, field managers, mobility planners and a chief administrative officer from an MPO.

In this research we follow up our technical analysis with in-depth interviews with representatives of California Metropolitan Planning Organizations (MPOs) and a survey of representatives of Transit Agencies that offer paratransit, about their interest in subcontracting some of their paratransit services to ride sharing companies or taxi services. There are 18 MPO's and 183 Transit Agencies in California. California presents a reasonable use case as the state includes areas with varied population densities and socioeconomic backgrounds. Though California has many large cities, there are many suburban and rural areas as well. We provide more detail about the effects of location on survey responses in section 5.3.

Note that our initial invitations via phone or e-mail led us to reduce the number of potential responding agencies to 157, as the others stated that they did not provide paratransit services. While we had hoped for a strong response rate, the rate of 19.1% is on the high side for external online surveys. Similarly to the synthesis study by the NSA. This work reflects the responses of 29 transit Agencies that responded. External surveys typically receive a 10-15% response rate (Nulty, 2008). See figure 1 for a diagram of our methodology.

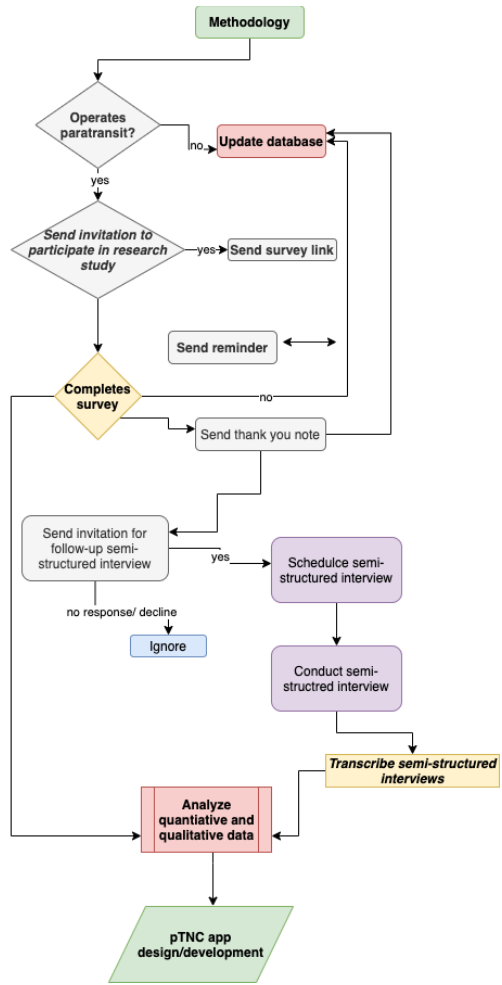

*Figure 1: Methodology Flowchart*

## 3 Literature Review

Our early work examined ways to develop IoT based blockchain enabled smart contracts to allow these extended paratransit systems. In the U.S., because of HIPAA (ADA (2009)), there are implied regulations which ensure the privacy of the disclosure of patron's personal health data. This established the first national standards in the United States to protect patients' personal or protected health information. The US department of Health and Human Services issued the rule to limit the use and disclosure of sensitive personal or protected health information in 1966. In theory, these systems must be privacy preserving and secure. The use of blockchain methods provides a way to securely create, store, and transfer digital assets in a distributed, decentralized environment. This is important when considering the incorporation of TNCs with paratransit operations.

However, synchronization in public blockchains typically requires significant computational power and extensive, if not enormous amounts of storage, which makes their use inefficient or infeasible for memory-limited IoT applications. To overcome this obstacle, our earlier research examined the use of a private blockchain for a paratransit system (Lewis and Regan, 2020). Private blockchains are permission based environments in which only the approved entities are able to access and add blocks. That initial work focuses on the design and development of working prototypes and examines the past and ongoing pilot programs scattered around the US. For example, the Massachusetts Bay Transportation Authority (MBTA) strived to improve the paratransit *RIDE* services with technological integration and collaboration with Uber and Lyft drivers to reduce service expenses for passengers, improve mobility management, and to provide on-demand individualized service using non-dedicated vehicles and reduce strain on the existing ADA programs (Kaufman et al., 2016).

The National Academies of Science (NAS) academies press published the TCRP project j-7 "Synthesis of information related to transit problems." the synthesis includes a literature review and profiles of 29 transit agencies that responded to the administrative survey and reflects various service models used for ADA (and coordinated) paratransit systems. The agencies were specifically chosen because of their geographic diversity, use of different service delivery models, and size. Of the 29 agencies, two California cities were represented, one in northern and one in southern California.

The study found that the most important benefits relayed by the 13 agencies that use taxis and other non-dedicated service providers (NDSPs) were the reduction in unit costs. This reduction resulted from the use of taxis to serve peak overflow and longer trips and to address the day-of-service need not offered by paratransit. However, the most prominent shortcomings include the degradation in service quality and not knowing the identity and location of a particular vehicle assigned. This issue was addressed in our survey as well. Also, there were security concerns that were posed as well stating that there was increased opportunities for fraud.

There is a section within the study specifically on "alternative service" Entitled *the use of taxis and transportation network companies for alternative services*. Of which, the authors define as TNC subsidy programs offered by a transit agency to ADA paratransit customers. Mentioning that: "While alternative services must be compliant with the ADA (in general), they are not governed by the service criteria of ADA complementary paratransit. Due to the following as projected by the authors:

1. The decision to use the alternative service is completely up to the customer.

2. While the transit agency can offer/suggest the alternative service option, the customer may still choose to use the ADA paratransit service.

3. A customer choosing to use the alternative service does not impact the customer's ADA paratransit eligibility or right to continue to request trips on the ADA paratransit service.

4. None of the vehicles used are owned, operated, or controlled by the transit agency.

Ultimately, trips served by these subsidy programs are not considered apart of the ADA paratransit service model and do not contribute to agencies meeting the ADA paratransit obligation. The study projects that the agencies paratransit service design model greatly influences the agencies decisions to offer alternative services. The study found that 41% of the surveyed agencies offer taxi-based alternative subsidy programs to their ADA paratransit customers. While we found that 41% of survey respondents have considered this partnership and about only 7% of our survey respondents conducted a pilot study.

## 4 ADA Paratransit Minimum Service Requirements

As briefly mentioned, paratransit services are obligated to meet some requirements. In this section, we have listed a few of the complementary paratransit services minimum requirements in terms of the service area, fare and response time.

- Service Area; must be within 3/4 miles on either side of a fixed route service.

- Fare; fares may not exceed twice the fare charge to an individual paying full fare for a fixed route trip of similar length, at similar time of day. Also, personal care attendant should not be charged any fee.

- Response time; must be provided at the requested time on particular day for service requests made the previous day. A call to the transit provider resulting in same day pickup is allowed but not mandated.

## 5 Research Design

The research design for this work consists of two stages of data collection and analysis. The first stage involved an online survey of transit agencies in the state of California, and the second stage involved semi-structured interviews with experts at Metropolitan Planning Organizations (MPOs) and transit agencies.

## 5.1 Survey Design

The surveys were completed online from early November 2019 to February 2020. The survey consisted of questions with a variety of response styles including: likert scale, true/false, demographic questions open and closed ended questions where respondents could elaborate on their responses. The short survey consisted of 10 questions, where the last question being a 5 part response. The responses are answers from the perspective on the transit agencies perceptions of the passenger experiences, their willingness to change and adapt to new technologies and willingness of agencies to provide adequate funding should paratransit ridership increase. Therefore, the responses may imply some biases. Major findings from the work can be found in sections 5 and 6.

## 5.2 Survey Respondent Demographics

The demographics of the respondents varied in different aspects. As depicted in figure 2, the type of agencies represented are as follows:

- Private-non profit corporation
- Private-for-profit corporation
- Independent public agency or authority of transit service
- City, county or local government unit or department of transportation
- Tribe
- Metropolitan Planning Organization (MPO), Council of Governments (COG) or other Planning Agency
- Other publicly owned or privately chartered corporation

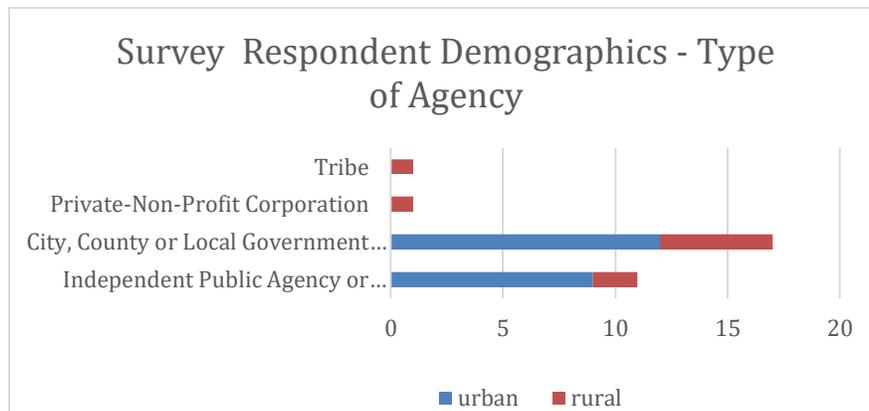

*Figure 2: Survey respondents categorized by the type of agency*

The majority of respondents were representative of city, county or local government units or departments of transportation. The survey responses are publicly available on GitHub [1] Each respondents' identity is protected and remains anonymous. The time to complete the survey was about 6 minutes.

About one third of the respondents service rural areas and the remaining service urban areas in California. The United States Department of Agriculture the Census Bureau provides the official, statistical definition of rural areas, based strictly on measures of population size and density. According to the current delineation, released in 2012 and based on the 2010 decennial census, rural areas comprise open country and settlements with fewer than 2,500 residents. The Census Bureau defines an urbanized area as one that includes an urban nucleus of 50,000 or more people. In general, they must have a core with a population density of 1,000 persons per square mile and may contain adjoining territory with at least 500 persons per square mile. Rural areas consist of open countryside with population densities less than 500 people per square mile and places with fewer than 2,500 people (Cromartie, 2019). We found that almost half of the survey respondents in rural areas reveal that their area is not serviced by TNCs like Uber and lyft.

---

[1] https://bit.ly/3bB4A1L

When asked- "Has your agency considered partnering with Taxi's, Uber or Lyft or other phone based services like Gogograndparent, which act as vetting services for TNC's like Uber and Lyft". We found that about that about 45% of paratransit agencies servicing rural populations have not considered such partnerships due to the lack of available TNC services in their service area. Additionally, 63% of paratransit agencies servicing urban areas have considered these partnerships and 17% of them have conducted a pilot.

**6 Survey Results –**

In this section, the survey responses are presented. *N* represents the total number of responses for that question.

|  | **Mean** | **N=** |
|---|---|---|
| Q1 - Approximately what percentage of paratransit passengers have smart phones? | 46% | 26 |
| Q2 - Approximately what percentage of paratransit passengers are over the age of 65? | 68% | 27 |
| Q3 - Approximately what percentage of paratransit passengers are under 25 years of age? | 14% | 27 |

Q4 - How do typical passengers request/schedule a ride with your paratransit services? 93% Call in advance. N=30

Q5. If passengers typically request/schedule rides by calling in advance, do passengers seem satisfied with the over the phone reservation service? – 93% yes, passengers seem satisfied with the over the phone call in advance service provided by paratransit. N=28

Q6. Has your agency considered partnering with Taxi's, Uber or Lyft or other phone based services like Gogograndparent, which act as vetting services for TNC's like Uber and Lyft?

43% Yes, we have considered this partnership 13% Yes, we have considered this partnership and conducted a pilot study 36% No, we have not considered this partnership 31.03% other- "Have discussed with other local agencies, but no action has been taken", No, not really available in rural area, partnered with X Taxi", No we do not have this service in our areas, No taxi, no Uber, No Lyft in the County, We partner with taxi's for a subsidized taxi program, Our county is not served by Uber, Lyft, and the like, For our regional service we have explored partnership but found options to be cost prohibitive given the long distance of trips, Concerned about ADA access and data reporting. N =30.

Q7. Has your agency made changes or improvements to paratransit services in the past decade? 80% yes, No 20%. Majority of improvements included the integration with some technology and the adoption of micro-transit. For example- "A same day service was added. This service allows customers to request trips on the day they want to travel. The customer pays the same cost of a trip for the first 5 miles. They would then pay the taxi rate for each additional mile after that. We then pay the taxi company a fixed per trip cost. Customers can also schedule their trips on the web and we are in the process of incorporating a service that will allow customers to pay for their trips online rather than pay the driver with cash or a coupon." N =30.

Q8. The use of any secure privacy preserving technology in communications in paratransit operations would add some cost to each ride, what is a reasonable price increase? 63% unsure. One respondent answering that the increase should remain >5% as "There is limited budget and the cost per ride for ADA is already astronomical." Another stating that a $0 increase "our area is extremely low income and cannot sustain a price increase." N=30.

Q9. Does your agency have a way to assess paratransit passenger satisfaction? Essentially, 83% answered yes with a variety of satisfaction methods including; customer satisfaction surveys, comment cards, complaints and more. 16% do not, one responded n/a. N=30.

Q10. Please indicate how you agree or disagree with the following statements related to the incorporation of private taxi's or ride-hailing vehicles.

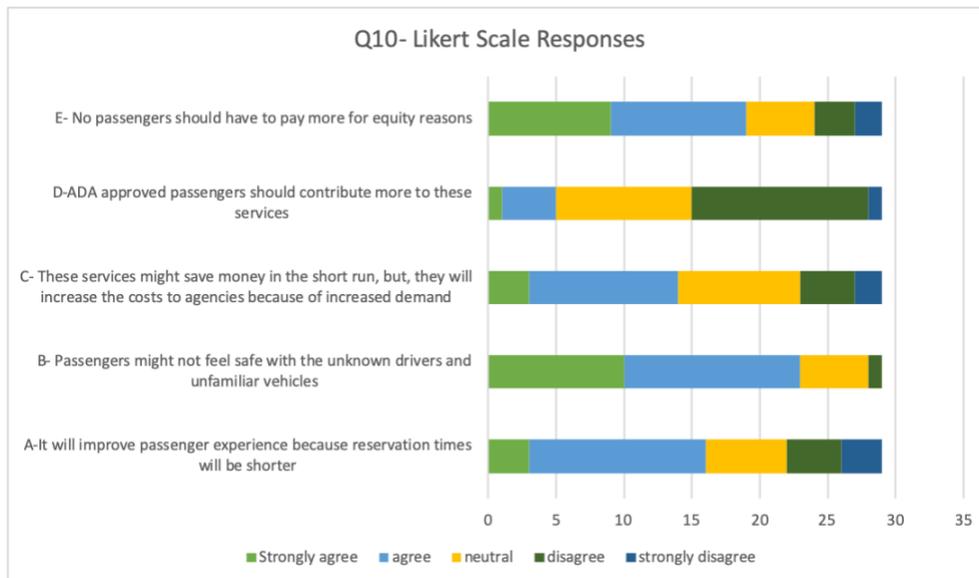

*Figure 4: Question 10 a-e survey response*

## 6.1 Analysis

We found that 90% of surveyed agencies reported that over half of their passengers are over the age of 65 years. To understand the passengers comfort levels with mobile technologies, and to see if there were any correlations between the passengers age group and smartphone usage we examined the correlation between the age responses and smart phone use, but did not find statistically significant correlations see figure 5.

In figure 4 below, the histogram presents the responses for participants on question 2 which asks the percentage of passengers over the age of 65. As we can see, majority reported that over half of their passengers are 65 years and older.

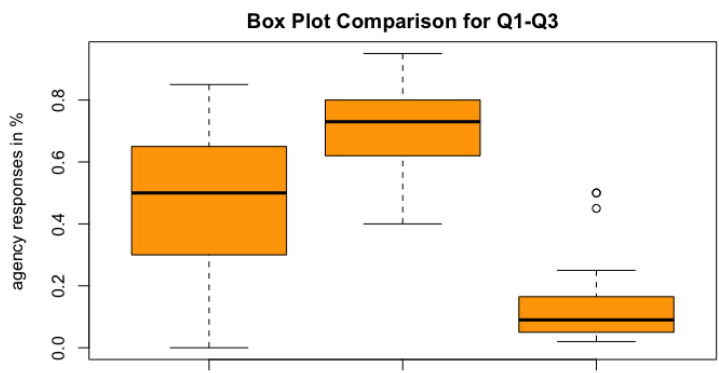

*Figure 5 box plot comparaison Q1-Q3*

## 7 Data Analysis - Mixed Methods

The analysis of our data incorporated a mixed methods approach using sequential explanatory design. This method is well suited for this work combining both quantitative and qualitative data. The quantitative data is primarily from the online survey and the qualitative data is from the semi-structured interviews with agencies and an MPO.

Securely encrypted software *Dedoose* was used to store and analyze the data, further analysis including correlations, regressions data and some visualizations were done in R studio. Results are found in the next sections.

### 7.1 Semi-structured Interviews

The semi-structured interviews consisted of roughly 15 questions. Since the interviews are semi-structured in nature, interviewees were able to elaborate on their responses and allude to more questions. The interviews were about an hour long and were all conducted at different date, times and locations. Each interview was conducted with two experts representatives. Participants; A-B, C-D, E-F. participant A - B; a Southern California MPO. Participant C - D two Northern California transit Agency categorized, as an independent public agency or authority of transit service. Participant E - F a Southern California transit Agency, categorized as a city government unit or department of transit. Through a weighted coding system, we were able to analyze the qualitative data from the interviews. The weighted scale from 1-5. 1-2 indicating a negative effect on paratransit operations based on the topic, 3 indicates neutral, and 4-5 indicating a positive effect. The themes are denoted as the codes for this analysis. Interviewing three transit agencies, one of which conduct a joint paratransit, and one MPO provided us with diverse perspectives on the interview questions. Through the interviews we have determined 14 major themes that were discussed by agencies and MPO interviewed and in figure 6 provides the code weighted statistics for each theme.

| Codes | Count | Min | Max | Mean | Median | Range | Sum | SD | Variance |
|---|---|---|---|---|---|---|---|---|---|
| **Aging Population** | 3 | 3 | 5 | 4 | 4 | 2 | 12 | 1 | 1 |
| **Autonomous Vehicles** | 2 | 3 | 4 | 3.5 | 3.5 | 1 | 7 | 0.7 | 0.5 |
| **Customer Feedback** | 12 | 2 | 5 | 4.6 | 5 | 3 | 55 | 0.9 | 0.8 |
| **Driver Shortage** | 3 | 1 | 3 | 2 | 2 | 2 | 6 | 1 | 1 |
| **Education** | 9 | 2 | 5 | 3.1 | 3 | 3 | 28 | 0.9 | 0.9 |
| **Environmental Impact** | 6 | 2 | 5 | 4.2 | 4.5 | 3 | 25 | 1.2 | 1.4 |
| **Funding** | 15 | 2 | 4 | 2.9 | 3 | 2 | 44 | 0.8 | 0.6 |
| **Government/Legislation** | 6 | 2 | 5 | 3.5 | 3.5 | 3 | 21 | 1.4 | 1.9 |
| **Growth/Expansion** | 9 | 1 | 5 | 2.8 | 3 | 4 | 25 | 1.5 | 2.2 |
| **Improvements** | 14 | 1 | 5 | 3.6 | 4 | 4 | 51 | 1.2 | 1.3 |
| **Operational Costs** | 7 | 1 | 5 | 2.6 | 2 | 4 | 18 | 1.4 | 2 |
| **Taxi or TNC pilot** | 20 | 1 | 5 | 2.8 | 3 | 4 | 55 | 1.3 | 1.6 |
| **Taxi/TNC lack of WAV** | 2 | 1 | 3 | 2 | 2 | 2 | 4 | 1.4 | 2 |
| **Technological advances** | 12 | 2 | 5 | 3.3 | 3 | 3 | 39 | 1 | 0.9 |

*Figure 6: Code weight statistics based on semi-structured interviews*

The codes were determined by the common themes raised during the interview through an axial coding process with a total of 120 excerpts. Using a weighted coding system, we were able to analyze the qualitative data from the interviews. The count indicates the number of times the concerns were raised by the interviewee. Additionally, in figure 7, the participant code frequency descriptor is available where participant A-B: 46 codes, participant C-D: 50 codes, participant E-F: 24 codes were extracted.

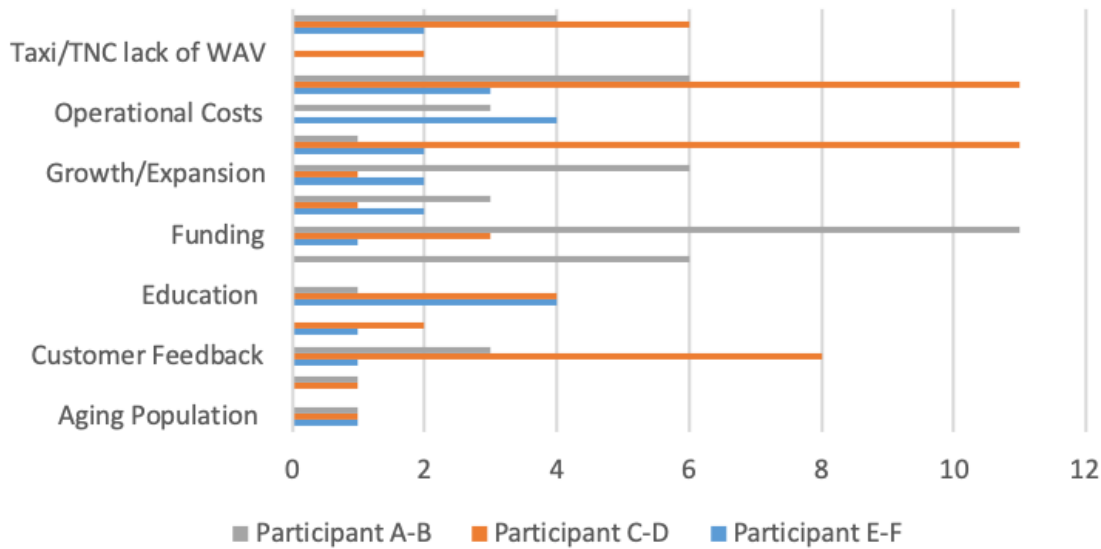

Figure 7: Code frequency descriptor

Major themes from the semi-structured interviews:

1. **Improvements** - Improvements to agencies are clear indications of progress and better passenger experiences. Though we found that about 31% of surveyed agencies have made improvements over the past decade. We followed up with this data by talking to agencies and gathering their insight about the improvements or lack thereof. In figure 7, we see that the majority of the discussion around improvements were done by participants C-D. Stating that: "...always pursuing improvements for the riders and their experiences...highest priority for us."

2. **Funding Concerns** - Funding was a concern of all agencies. As presented in figure 4 we can see how prevalent funding concerns are. Enhanced Mobility of Seniors and Individuals with Disabilities - Section 5310 of the Federal Transit Administration's mission and focus is to innovate by developing and supporting transit programs and services in rural and small-city America. The 5310 funding is provided to states for the purpose of assisting private nonprofit groups in meeting the transportation needs of older adults and people with disabilities when the transportation service provided is unavailable, insufficient or inappropriate to meeting these needs.

   The Transportation Development Agency (TDA) provides funding to counties for transit and non-transit related purposes. In the synthesis study, the authors mention that there is increasing pressure on transit agencies to be as cost efficient as possible while maintaining service quality standards. That was a major theme during the interviews with Agencies. The funding concerns mostly negatively impacted agencies.

   Eligible projects include both "traditional" capital investment and "nontraditional" investment beyond the Americans with Disabilities Act (ADA) complementary paratransit services.

3. **Education** - Education was also an important element that interviewees mentioned. educating stakeholders, government officials, and passengers on important aspects of transportation. though transportation is a complex field. It is continually growing and developing as society progresses. With the adoption of new technologies and the implementation of new policies it is important to educate others on the progress and changes.

4. **Technological advances and challenges** - Lastly, paratransit operators in the northern regions of California desire better software for their systems, stating "Big challenges is educating people and getting the technology right...only a few software manufactures on the market for paratransit, it is a very niche market." and "It would be nice if the companies that make the software that we rely on everyday had more of an understanding of how that software should be used."

5. **TNCs lack of WAVs** - The lack of accessible vehicles available by taxis and TNCs is a major concern of paratransit operators. Due to the low number of these vehicles, operators do not want to take the risk. The primary goal of paratransit is to provide accessibility. With the adoption of these services there would be a separation of wheelchair users and those with other mobility, vision, or cognitive disabilities. Although we did not receive a high rate of conversation around this topic during the interviews, our survey data supports this theme as we had more representation from rural area.

   According to our survey, an average of 23% of passengers require WAVs, many of these passengers being 65 years and over. Participant E-F noted an increase in the re-certifications for paratransit but a decrease in new certifications. Those respondents also projected an increase in the demand for paratransit in the future, suggesting that a supply-demand imbalance is on the horizon. Participants E's agency does offer a same day-taxi service to its passengers of which 46% of active users and 8% of their overall eligible customers provides more flexibility to their passengers and this is also at a cost saving to the agency. participant E-F's agency has successfully sub-contracted with a local company operating sedans and the sub-contractor is responsible for 30-40% of their trips.

**8 Conclusions and Future work**

Agencies and MPOs are monitoring and forecasting progress in paratransit. Within our study, we gained a sense of the sentiments towards paratransit and its future. The small dataset containing the 30 responses yields to interesting insight. Through this work we were able to establish relationships with the agencies in California. On the planning side, MPOs are in charge of ensuring that paratransit agencies and operators are compliant with the ADA as well as the government initiatives for environmental sustainability and the reduction of greenhouse gas emissions. Conversely, on the agency side, the focus is always to ensure that the passengers are provided adequate service and satisfaction. Paratransit agencies are obligated and required to provide ADA compliant services, however; we found that they are proud to offer additional services to better accommodate the diverse transportation services to meet the needs of the diverse range of passengers, and essentially, these services are more cost efficient than operating paratransit and service ADA and non ADA passengers. The agencies' funding plays a major role in their services. There are several funding sources for agencies in California including; 5310 and TDA.

Ubers, Lyft taxi's and other TNCs are using secure technologies to provide on-demand services. But, those service providers can be unreliable, and that would negatively impact the sub-contracting experience. One of the agencies that we interviewed were negatively impacted from sub-contracting with a local taxi company. Initially, the collaboration was positively received but, the agency experienced the sub-contracting taxi drivers cancellations, delays and in the end, the company went out of business on very short notice and left the agency hours to fulfil the unattended rides. Though we are presenting anecdotal information here, we believe that the experience described is not uncommon.

Lastly, we cannot resolve the location challenges of TNCs but, we do believe that this partnership would encourage expansions of service areas. We found that a high percentage of our rural surveyed population did not have access to these TNCs.

We conclude that the adoption of secure technologies may be achieved internally to ensure reliability, security, and passenger satisfaction. The agencies response to the adoption of these sub-contracted services were varied. Many have already outsourced to TNCs. The major concern for paratransit agencies is they all want to provide satisfaction to customers and they are all mandated to at least provide service that complies to the ADA's minimum requirements.

We believe that through a cohesive collaborative effort among TNCs and paratransit operators that customer services can be improved and agency costs contained.

In the next phase of this work, we plan to conduct a focus groups and a survey of paratransit passengers at a to understand and document their experiences and perspectives on paratransit and the adoption of TNC services.

## 9 Acknowledgement

This research is supported by The National Science Foundation Graduate Research Fellowship Program and the Pacific Southwest Region Transportation Center. We would like to thank all of the agencies that participated in this research and support this work.

## References

ADA. 2009. *Americans with Disabilities Act*. https://www.ada.gov/cguide.htm.

Cromartie, John. 2019. *Data for Rural Analysis*. https://www.ers.usda.gov/topics/rural-economy-population/rural-classifications/what-is-rural/.

Forster, P. W., & Regan, A. C. (2001). Electronic integration in the air cargo industry: An information processing model of on-time performance. *Transportation Journal*, 46-61.

Gogograndparent. 2019. *GoGoGrandparent:Transportation for Seniors*. https://bit.ly/31YqAxA.

Health, US, and Human Services. 2019. *Hippa Compliance Rules*. https://www.hhs.gov/hipaa/for-professionals/privacy/index.html.

Javaid, U., Aman, M. N., & Sikdar, B. (2019, April). DrivMan: Driving trust management and data sharing in VANETS with blockchain and smart contracts. In *2019 IEEE 89th Vehicular Technology Conference (VTC2019-Spring)* (pp. 1-5). IEEE.

Kanza, Y., & Safra, E. (2018, November). Cryptotransport: blockchain-powered ride hailing while preserving privacy, pseudonymity and trust. In *Proceedings of the 26th ACM SIGSPATIAL International Conference on Advances in Geographic Information Systems* (pp. 540-543).

Kaufman, S. M., Smith, A., O'Connell, J., & Marulli, D. (2016). Intelligent Paratransit.

Lewis, Amari, and Amelia Regan. 2020. "Enabling Paratransit and Tnc Services with Blockchain Based Smart Contracts." In *Proceedings of the Computing Conference 2020 (in Press)*. SAI (The Science; Information Organization).

Luo, Y., Jia, X., Fu, S., & Xu, M. (2018). pRide: Privacy-preserving ride matching over road networks for online ride-hailing service. *IEEE Transactions on Information Forensics and Security*, *14*(7), 1791-1802.

Lyft. 2020. *Lyft Business*. https://www.lyftbusiness.com/industries/healthcare.

Nulty, D. D. (2008). The adequacy of response rates to online and paper surveys: what can be done?. *Assessment & evaluation in higher education*, *33*(3), 301-314.

Uber. 2020. *Uber Assist*. https://www.uber.com/blog/los-angeles/introducing-uberassist-la/.

Uber WAV. 2020. *WAV*. https://www.uber.com/us/en/ride/uberwav/.